\documentstyle[twocolumn,aps,epsfig,floats]{revtex}
\begin{document}

\twocolumn[\hsize\textwidth\columnwidth\hsize\csname %
@twocolumnfalse\endcsname

\title{Magnetic Coherence in Cuprate Superconductors}
\author{Dirk K. Morr $^{1,2}$ and David Pines $^{1,3}$}
\address{ $^{1}$ University of Illinois at Urbana-Champaign,
Loomis Laboratory of Physics, 1110 W. Green St., Urbana, IL
61801\\
$^{2}$ Theoretical Division, Los Alamos National
Laboratory, Los Alamos, NM 87545\\
$^{3}$ Institute for Complex Adaptive Matter,
University of California, and LANSCE-Division, Los Alamos National
Laboratory, Los Alamos, NM 87545}
\date{\today}
\draft
\maketitle
\begin{abstract}
Recent inelastic neutron scattering (INS) experiments on
La$_{2-x}$Sr$_x$CuO$_4$ observed a {\it magnetic coherence effect}, i.e.,
strong frequency and
momentum dependent changes of the spin susceptibility,
$\chi''$, in the
superconducting
phase.
We show that this
effect is a direct consequence of changes in
the damping of incommensurate antiferromagnetic spin fluctuations
due to the appearance of a d-wave gap in the fermionic spectrum.
Our theoretical results provide a quantitative
explanation for the weak momentum dependence of the observed spin-gap. Moreover,
we
predict {\bf (a)} a Fermi surface in La$_{2-x}$Sr$_x$CuO$_4$ which is
closed around $(\pi,\pi)$ up to optimal doping, and {\bf (b)} similar changes in
$\chi''$ for all cuprates with an
incommensurate magnetic response.

\end{abstract}
]

\narrowtext

The spin excitation spectrum in La$_{2-x}$Sr$_x$CuO$_4$
\cite{Shi89,Mas96,Aep97,Yam98,Lake99} in the normal and
superconducting state has been intensively studied during the last
few years in inelastic neutron scattering (INS) experiments. The
normal state spectrum for  compounds with $x>0.04$ is
characterized by peaks in $\chi''({\bf q}, \omega)$ at
incommensurate wave-vectors ${\bf Q}_i=(1 \pm \delta,1) \pi$ and
${\bf Q}_i=(1,1 \pm \delta) \pi$ \cite{Shi89,Mas96,Yam98}, where
$\delta$ increases with increasing doping. Recent INS experiments
in the superconducting state of La$_{2-x}$Sr$_{x}$CuO$_4$ (LSCO)
by Mason {\it et al.}~(x=0.14) \cite{Mas96} and
Lake {\it et al.}~(x=0.16) \cite{Lake99} show striking momentum and frequency
dependent changes in $\chi''$ upon entering the superconducting
state, which the authors called the {\it magnetic coherence effect}. For
both compounds, $\chi''({\bf Q}_i)$ in the superconducting state
is considerably decreased from its normal state value below
$\omega \approx 7$ meV, while it increases above this frequency.
For frequencies in the vicinity of 7 meV, the incommensurate peaks
sharpen in the superconducting state,  while at higher frequencies
the peak widths in the normal and superconducting state are
approximately equal. Moreover, by employing a Kramers-Kronig
transformation, the authors found that the static susceptibility,
$\chi'$,
at ${\bf Q}_i$ decreases in the superconducting state
\cite{Lake99}.

In this communication, we show that the
magnetic coherence effect is a direct consequence of changes in
the damping of incommensurate antiferromagnetic spin fluctuations
due to the appearance of a d-wave gap in the fermionic spectrum.
We obtain results for the frequency and
momentum dependence of $\chi''$ that are in good qualitative,
and to a large extent quantitative agreement with the experimental
data, and also explain the weak momentum
dependence of the spin-gap. We show that INS data in the
superconducting state provide information on the
symmetry of the order parameter and the topology of the Fermi surface (FS) and
that for La$_{2-x}$Sr$_x$CuO$_4$, INS experiments suggest a
FS closed around $(\pi,\pi)$. We predict that the
magnetic coherence effect is to be expected for any cuprate
superconductor with an incommensurate spin spectrum, and thus in
particular for YBa$_2$Cu$_3$O$_{6+x}$ (YBCO), in which an
incommensurate spin structure at low frequencies has been observed
\cite{Tra92,Dai97,Mook98}.

The starting point for our calculations is a spin-fermion
model \cite{sfmodel} in which the damping of incommensurate
spin-excitations arises from their interaction with fermionic
quasi-particles. In this model, the spin propagator,
$\chi$, is given by
\begin{equation}
\chi^{-1} = \chi_0^{-1} -  \Pi  \ ,
\label{Dyson}
\end{equation}
where $\chi_0$ is the bare propagator, and $\Pi$ is the bosonic
self-energy given by the irreducible particle-hole bubble. $\chi_0$ is in
general obtained by integrating
out the
high-energy fermionic degrees of freedom. However, since the form of
fermionic excitations at high frequencies is so far not well understood,
a microscopic calculation of $\chi_0$ is not yet feasible.
We therefore make the experimentally
motivated ansatz
\begin{equation}
\chi_0^{-1}= { \xi_0^{-2} + ({\bf q} - {\bf Q}_i)^2 \over
\alpha } \ ,
\label{chi0}
\end{equation}
where $\xi_0$ is defined as the ``bare"  magnetic correlation length
(unrenormalized by the coupling to low-frequency particle-hole excitations) and
$\alpha$ is a temperature independent constant. In general one
would expect a frequency term in Eq.(\ref{chi0}) which is omitted
here because experimentally there is no observed dispersion in the spin
excitation spectrum below
$\omega \approx 25$ meV \cite{Maspc}, well above the frequency
range we consider here. The above form of $\chi_0$ thus {\it only}
determines the position of the incommensurate peaks in momentum
space; it does {\it not} affect the frequency dependence of
$\chi''$, which arises solely from $\Pi$. In the
following we define the renormalized magnetic
correlation length as $\xi^{-2} = \xi_0^{-2} - \alpha \, {\rm Re} \, \Pi$.

We first consider $\chi''$ at ${\bf Q}_i$ in the normal state.
Calculating $\Pi_N$ to lowest order in the spin-fermion coupling
$g$ yields Im$\, \Pi_{N} \sim \omega$, while Re$\, \Pi_{N}
\approx$ const.~\cite{Morr99a}, yielding $\xi_{N}^{-2}(\omega) =
const$, and a frequency dependent
dynamic susceptibility, $\chi''({\bf Q}_i, \omega)$, of the MMP
form \cite{MMP} which quantitatively describes the results of INS
experiments in the normal state of LSCO \cite{MMP} and YBCO
\cite{Morr99}.

In the superconducting state, $\Pi_{SC}$ is given by (to lowest order in $g$)
\begin{eqnarray}
\Pi_{SC}({\bf q}, i \omega_n) &=& -g^2 \, T \sum_{{\bf k},m} \ \Big\{
G({\bf k}, i\Omega_m) G({\bf k+q}, i\Omega_m+i\omega_n) \nonumber
\\ & & + F({\bf k}, i\Omega_m) F({\bf k+q}, i\Omega_m+i\omega_n)
\Big\} \ , \label{Pi}
\end{eqnarray}
where $G$ and $F$ are the normal and anomalous Green's functions
\begin{eqnarray}
G&=&{ i\omega_n + \epsilon_k \over (i \omega_n)^2 -\epsilon_k^2 -
\Delta_k^2 }, \ \ F={\Delta_k \over (i \omega_n)^2 -\epsilon_k^2 -
\Delta_k^2 } \ .
\end{eqnarray}
$E_{\bf k}= \sqrt{ \epsilon_{\bf k}^2 + |\Delta_{\bf k}|^2}$ is
the fermionic dispersion in the super\-conduc\-ting \- state, \-
$\Delta_{\bf k}=\Delta_{0}\left( \cos(k_x) - \cos(k_y) \right)/2$
is the d-wave gap and
\begin{eqnarray}
\epsilon_{\bf k} &=& -2t \Big( \cos(k_x) + \cos(k_y) \Big) \nonumber \\
& & \quad -4t^\prime \cos(k_x) \cos(k_y)  -\mu \ ,
\label{dispersion}
\end{eqnarray}
is the electronic tight-binding dispersion where $t, t^\prime$ are
the hopping elements between nearest and next-nearest neighbors,
respectively, and $\mu$ is the chemical potential. Since our
theoretical results for both doping levels of LSCO $x=0.14(0.16)$
are quantitatively similar, we consider for definiteness
$x=0.16$ and choose $t^\prime/t=-0.22$ and $\mu/t=-0.84$, a choice
which will be
seen to yield agreement with the experimental data. The
superconducting gap, $\Delta_{0}\approx 10$ meV, is taken to be that extracted
from Raman scattering experiments by Chen {\it et
al.}\cite{Chen94} and the incommensurate wave-vector ${\bf Q}_i$
is at $\delta\approx 0.25$ \cite{Mas96,Yam98}.
\begin{figure} [t]
\begin{center}
\leavevmode
\epsfxsize=7.5cm
\epsffile{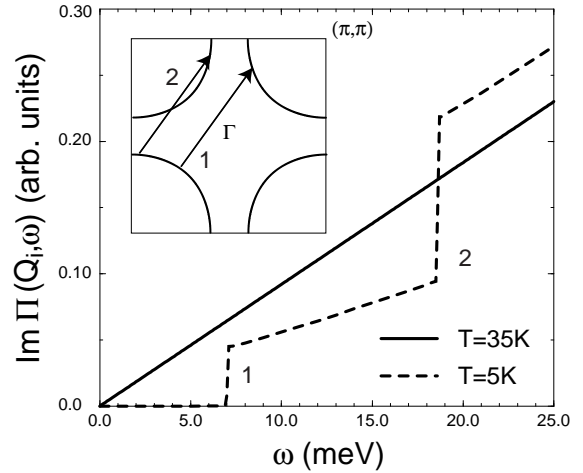}
\end{center}
\caption{ The spin-damping Im$\, \Pi$ at ${\bf Q}_i$ as a function
of frequency in the normal (solid line) and superconducting state
(dashed line). Inset: Fermi surface of La$_{2-x}$Sr$_x$CuO$_4$ and
quasiparticle threshold transitions with wave-vector ${\bf Q}_i$.}
\label{sd_om}
\end{figure}

Our theoretical results for the spin-damping,
Im$\, \Pi$ in Eq.(\ref{Pi}) at ${\bf Q}_i$ are presented in
Fig.~\ref{sd_om}. Since ${\bf Q}_i$ is incommensurate, we obtain
four decay channels for spin excitations; the two channels in the
first Brillouin zone are shown in the inset of Fig.~\ref{sd_om}.
In the normal state all four channels for particle-hole
excitations are excited in the low frequency limit, which
yields Im$\, \Pi_N \sim \omega$, as noted above. In the
superconducting state, the four channels split into two pairs with
degenerate non-zero threshold energies, $\omega_c^{(1,2)}$, that are
determined by the
momentum dependence of the order parameter and the shape of the
Fermi surface. In particular, we find
$\omega_c^{(1,2)}=|\Delta_{\bf k}|+|\Delta_{\bf k+Q_i}|$, where ${\bf k}$ and
${\bf k+Q}_i$ both lie on the Fermi surface, as shown in the inset of
Fig.~\ref{sd_om}. For the band parameters chosen, the threshold
energies are $\omega_c^{(1)}=0.70 \Delta_{SC}$ for quasiparticle
excitations close to the nodes of the superconducting gap
(excitation 1), and $\omega_c^{(2)}=1.86 \Delta_{SC}$ for
excitations which connect momenta around $(0,\pi)$ and $(\pi,0)$
(excitation 2).  Note that due to the superconducting coherence
factors in Eq.(\ref{Pi}), Im$\, \Pi_{SC}$ exhibits sharp jumps at the
threshold frequencies. Since for $T=0$ and $\omega<\omega_c^{(1)}$, Im$\,
\Pi_{SC} \equiv 0$, and thus $\chi_{SC}'' \equiv 0$,
$\omega_c^{(1)}$ is often referred to as the spin-gap in the
superconducting state.

We now turn to the calculation of Re$\, \Pi_{SC}$. The gap in
Im$\, \Pi_{SC}$ gives rise to a $\omega^2$-term in Re$\, \Pi_{SC}$
for $\omega \ll \omega_c^{(1)}$, while Re$\, \Pi_{SC} \approx
const.$ for $\omega \gg \omega_c^{(2)}$. The steps in
Im$\, \Pi_{SC}$ at $\omega_c^{(1,2)}$ create logarithmic divergences in
Re$\, \Pi_{SC}$; as has recently been demonstrated \cite{Chu98}
these are an artifact of our restriction to the second order
bosonic self-energy correction. When fermionic lifetimes are
calculated within a self-consistent strong-coupling approach, the
authors of Ref.\cite{Chu98} found that the steps in
Im$\, \Pi_{SC}$ are smoothed out, while the gap below a frequency
$\approx \omega_c^{(1)}$ still persists. The weak logarithmic
divergences in Re$\, \Pi_{SC}$ become a smooth function of
frequency. Since the spin-gap survives the inclusion of realistic
fermionic lifetimes, we expect the conclusions we draw in the
following to be valid beyond the current level of approximation.

\begin{figure} [t]
\begin{center}
\leavevmode
\epsfxsize=7.5cm
\epsffile{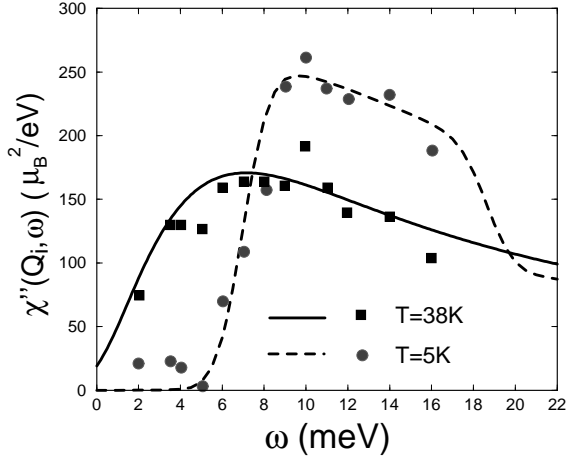}
\end{center}
\caption{ Fit of our theoretical results for $\chi''({\bf Q}_i,
\omega)$ to the experimental data Ref.\protect\cite{Lake99}
(filled circles and squares) in the
normal (solid line) and superconducting
state
(dashed line).}
\label{chi_om}
\end{figure}

In Fig.~\ref{chi_om} we present a fit of our theoretical results
for $\chi''$ to the experimental data of Ref.~\cite{Lake99} in the
normal and superconducting state.
The fit to the experimental data in the
superconducting state for $\omega_c^{(1)}<\omega<16$ meV was
obtained by making the ansatz that at these frequencies, $\xi_{SC}(\omega)$ is
frequency independent and given by
\begin{equation}
\xi^{2}_{SC}(\omega) \approx  { 3 \over 2} \, \xi^{2}_N = const.
\label{xi_sc}
\end{equation}
To account for the experimental energy resolution we convolute our
theoretical results with a Gaussian distribution of width $\sigma
\approx 2$ meV.

The ``high frequency" result, Eq.(\ref{xi_sc}), is a consequence of the
redistribution in the spectral weight of $\chi''({\bf Q}_i, \omega)$ in the
superconducting state. By using the Kramers-Kronig relation, we find
$\xi_{SC}(\omega=0)<\xi_{N}(\omega=0)$ while, as seen in Fig.~\ref{chi_om} quite
generally, for $\omega_c^{(1)}<\omega<\omega_c^{(2)}$, $\chi_{SC}''(\omega)$
exceeds $\chi_{N}''(\omega)$, so that $\xi_{SC}(\omega)>\xi_{N}(\omega)$ in this
frequency range. As may be seen in Fig.~\ref{chi_om}, the simple ansatz,
Eq.(\ref{xi_sc}), yields good agreement with experiment. Because the form of
both $\chi_{SC}''(\omega)$ and $\chi_{N}''(\omega)$ above $\omega_c^{(2)}$ is
not well known, one cannot at present arrive at a self-consistent description of
the frequency dependence of $\xi_{SC}$ using the Kramers-Kronig relation. We
note, however, that upon restricting the frequency integration in the
Kramers-Kronig relation to $\omega<\omega_c^{(2)}$, we find
$\chi_{SC}'(\omega=0)
\approx 0.65 \chi_{N}'(\omega=0)$, in agreement with the results of
Ref.~\cite{Lake99}.

\begin{figure} [t]
\begin{center}
\leavevmode
\epsfxsize=7.5cm
\epsffile{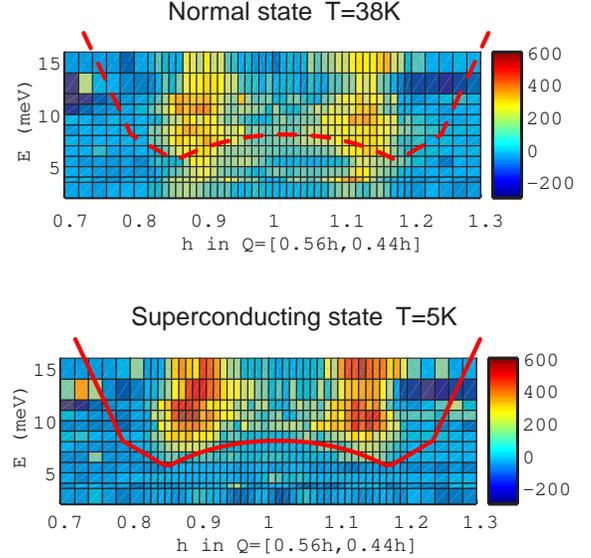}
\end{center}
\caption{Experimental intensity in the $(\omega,{\bf q} )$-plane
for {\it (a)} the normal and {\it (b)} superconducting state,
together with our theoretical results for $\omega_c^{(1)}({\bf
q})$ (red line). The coloring of the squares indicates the
intensity. Experimental data are taken from
Ref.~\protect\cite{Lake99}.}
\label{spingap}
\end{figure}

\begin{figure} [t]
\begin{center}
\leavevmode
\epsfxsize=7.5cm
\epsffile{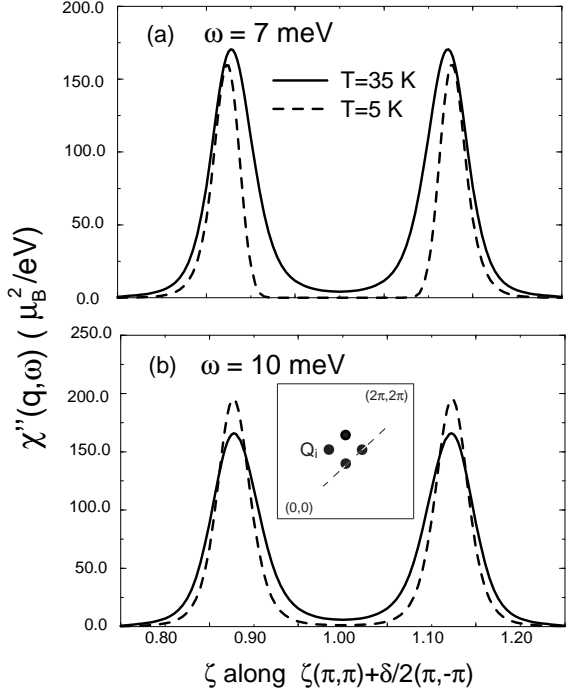}
\end{center}
\caption{  $\chi''({\bf q}, \omega)$ in the normal (solid line)
and superconducting state (dashed line) for {\it (a)} $\omega=7$
meV  and {\it (b)} $\omega=10.0$ meV, along the momentum space
path shown in the inset of {\it (b)}.}
\label{chi_q}
\end{figure}

To understand the momentum dependent changes in $\chi''$
between the normal and superconducting state, we
consider the momentum dependence of the spin-gap,
$\omega_c^{(1)}({\bf q})$. In Fig.~\ref{spingap} we plot the
experimental intensity in the $(\omega,{\bf q} )$-plane for the
momentum space path shown in the inset of Fig.~\ref{chi_q}b,
together with our theoretical results for $\omega_c^{(1)}({\bf
q})$ (red line).
We also included $\omega_c^{(1)}({\bf q})$ as a dashed red line
into the normal state data so as to demonstrate the transfer of
spectral weight between the normal and the superconducting state
from frequencies below $\omega_c^{(1)}({\bf q})$ to frequencies
above the spin-gap. A comparison of Fig.~\ref{spingap}a and b
clearly shows that the experimental intensity that exists in the
normal state for $\omega<\omega_c^{(1)}({\bf q})$ vanishes as
expected in the superconducting state.
Note that the momentum dependence of the spin-gap is rather weak;
it only changes from $\Delta_{sg}^{min} \approx 5.5$ meV at its
minimum (close to ${\bf Q}_i$) to $\Delta_{sg}^{max}\approx 8$ meV
at its local maximum midway between the incommensurate positions.
We thus conclude that our theoretical result for the momentum
dependence of the spin gap
provides a good quantitative description of the area in the
$(\omega,{\bf q} )$-plane where the spectral weight vanishes in
the superconducting state.

We now turn to momentum dependence of $\chi''$ in the normal and
superconducting state. Our theoretical results, which we present
in Fig.~\ref{chi_q}, correspond to horizontal cuts in the
$(\omega,{\bf q} )$-plane of Fig.~\ref{spingap} at $\omega=7$ and
10 meV.
For $\omega=7$ meV (Fig.~\ref{chi_q}a), the peak intensity in the
superconducting state is anisotropically reduced, with a stronger
suppression of $\chi_{SC}''$ towards the center of the scan. The
anisotropic suppression of $\chi_{SC}''$ is a direct consequence of
the momentum dependence of the spin-gap which increases when moving
from ${\bf Q}_i$ towards the center of the scan, but decreases in
the opposite direction. As a result,
$\chi_{SC}''$ is rapidly cut off by the spin-gap when moving
towards the center of the scan, but is scarcely reduced in the opposite
direction. This peak anisotropy should be observable for all frequencies
between
$\Delta_{sg}^{min}$ and $\Delta_{sg}^{max}$. For $\omega=10$ meV
$>\Delta_{sg}^{max}$ (Fig.~\ref{chi_q}b), the anisotropy
vanishes, and the peak intensity increases in the superconducting
state, as expected from Fig.~\ref{chi_om}. Since the anisotropy of
$\chi_{SC}''$ around ${\bf Q}_i$ is reduced with increasing
frequency, the peak maximum seems to slightly shift towards the
center of the scan. All these results, i.e., narrow peaks around
7-8 meV, a simultaneous increase in peak width and height and a
shift of the peak maximum towards the center of the momentum scan
with increasing frequency agree qualitatively with the
experimental findings of Ref.~\cite{Mas96,Lake99} as may be seen by comparing
Fig.~\ref{chi_q}a (Fig.~\ref{chi_q}b) with the
bottom (top) part of Fig.~2 in Ref.~\cite{Mas96} or with Fig.~2b (Fig.~2c) in
Ref.~\cite{Lake99}. Mason {\it
et al.}~considered the sharpening of the incommensurate
peaks in the superconducting state as an effect of {\it magnetic
coherence} which is shown here to arise solely from the momentum
dependence of the spin-gap.

Due to the symmetry of the Fermi surface, $\chi_{SC}''$ exhibits four different
threshold frequencies for
momenta away from ${\bf Q}_i$. While the two upper
thresholds
remain close, the energy separation between the two lower
threshold
increases rapidly with distance from ${\bf Q}_i $
\cite{Morr99a}. We thus predict that $\chi_{SC}''$ will acquire additional
frequency
structure for ${\bf q} \not = {\bf Q}_i$.

The INS data also provide insight into the form of the FS in LSCO
and thus complement the results of angle-resolved photoemission
(ARPES) experiments. Since excitation (1) is located in the
vicinity of the nodes where the superconducting gap changes
rapidly with momentum, $\omega_c^{(1)}$ sensitively depends on the
form of the FS and the symmetry of the
order parameter. The frequency location of $\omega_c^{(1)}$ can
therefore be used to extract information on the form of the Fermi
surface, and, within the framework of Eq.(\ref{dispersion}), on
the value of $t^\prime/t$. In particular, we find that the INS
data provide a lower bound for $t^\prime/t$. Within our scenario,
excitation (1) across the FS (see inset of Fig.~\ref{sd_om})
becomes impossible in the superconducting state for $|t'|<0.2t$.
Since this implies $\chi''({\bf Q}_i)=0$ for frequencies below
$\omega_c^{(2)}$, in contradiction to the experimental results, we
conclude $|t'| \geq 0.2t$. Assuming a weak doping dependence of
$t^\prime/t$, this constraint for $t'$ yields a FS of LSCO which
is closed around $(\pi,\pi)$ up to optimal doping. The FS thus
possesses the same topology as that in YBCO; this explains the occurrence
of incommensurate peaks in
the spin spectrum along the same direction in momentum space in
the latter materials \cite{Tra92,Dai97,Mook98}.

Though the details of the magnetic coherence effect are sensitive
to material specific parameters, e.g., Fermi surface topology, the
extent of the incommensuration $\delta$, their experimental
observation  only depends on two criteria:
the existence of an incommensurate spin structure and the d-wave
symmetry of the superconducting gap. We thus predict a similar
effect for all cuprate superconductors in which these criteria are
met, and are currently studying its form in YBCO \cite{Morr99a}.

Finally, the theoretical scenario for the  magnetic coherence
effect presented here is conceptually different from that
recently proposed for the {\it resonance peak}
\cite{Morr98}. Not only do the effects take place in different
wave-vector and energy regions of $\chi''$ \cite{Mook98}, but the
origin of the {\it resonance peak} is ascribed to a dispersing
spin mode, while our scenario for the coherence effect is solely
based on the existence of a relaxational spin mode.

In summary, we find that the frequency and momentum dependent
changes of $\chi''$ in the superconducting state are a direct
consequence of changes in the quasiparticle spectrum due to the
appearance of a d-wave gap. We show that the available INS data
constrain the Fermi surface topology, and suggest a Fermi surface
in La$_{2-x}$Sr$_x$CuO$_4$ which is closed around $(\pi,\pi)$ up
to optimal doping. We make several predictions for the frequency
dependence of $\chi''(\omega)$ at and around ${\bf Q}_i$ which
await further experimental testing. Finally, we predict the
presence of comparable changes in $\chi''$ in all cuprate
superconductors with an incommensurate spin-structure.

We would like to thank G. Aeppli, A.V. Chubukov, P. Dai, A.
Millis, H. Mook, and J. Schmalian for valuable discussions and
particularly B. Lake and T. Mason for very stimulating discussions
and for providing us with their experimental data prior to
publication. This work has been supported in part by the Science
and Technology Center for Superconductivity through NSF-grant
DMR91-20000, and by DOE at Los Alamos.

\end{document}